\documentclass{jaa}
\usepackage{natbib}
\usepackage{graphicx}

\usepackage[section]{placeins}
\usepackage{subfig}
\usepackage{graphicx}
\usepackage{lipsum}
\usepackage{footnote}
\usepackage{xcolor}
\usepackage{hyperref} 
\hypersetup{allbordercolors=white}
\usepackage{rotating}

\newcommand{\orcid}[1]{\href{https://orcid.org/#1}{\includegraphics[width=10pt]{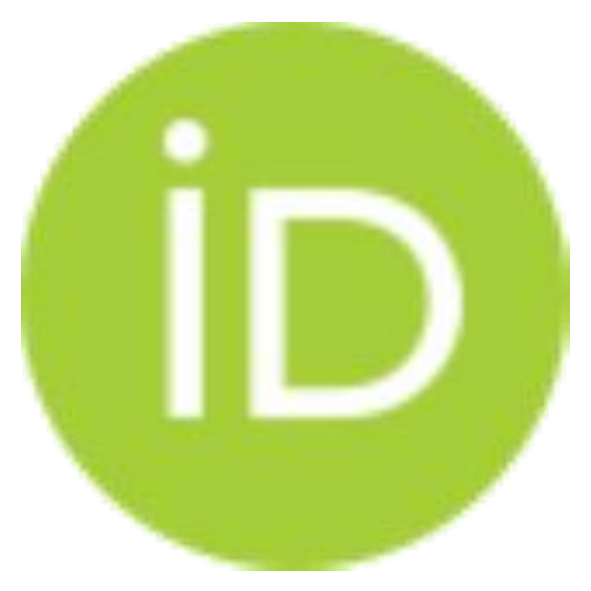}}}
\newcommand{\SMMFphotosphere}{$B_{\mathrm{p}}$\,}
\newcommand{\SMMFcore}{$B_{\mathrm{c}}$\,}
\newcommand{\SMMFwings}{$B_{\mathrm{w}}$\,}
\newcommand{\SMMFhmi}{$B_{\mathrm{hmi}}$\,}
\newcommand{\SMMFwso}{$B_{\mathrm{wso}}$\,}

\DeclareRobustCommand{\ion}[2]{\textup{#1\,\textsc{\lowercase{#2}}}}
\newcommand{\FeIline}{\ion{Fe}{i}~6301.5\,\AA\,}
\newcommand{\CaIIline}{\ion{Ca}{ii}~8542\,\AA\,}
\newcommand{\wsoFeIline}{\ion{Fe}{i}~5250\,\AA\,}
\newcommand{\hmiFeIline}{\ion{Fe}{i}~6173\,\AA\,}

\begin{document}\sloppy

\title{Solar Mean Magnetic Field of the Chromosphere}

\author{M. Vishnu\,\orcid{0000-0002-8942-6774}\,\textsuperscript{1,*},
        K. Nagaraju\,\orcid{0000-0002-0465-8032}\,\textsuperscript{1},
        Harsh Mathur\,\orcid{0000-0001-5253-4213}\,\textsuperscript{1}}  


\affilOne{\textsuperscript{1}Indian Institute of Astrophysics, Koramangala 2$^{nd}$ Block, Bengaluru, PIN - 560034, India\\}

\twocolumn[{
\maketitle

\corres{vishnu.madhu@iiap.res.in}
\msinfo{** *** ****}{** *** ****}


\begin{abstract}

The Solar Mean Magnetic Field (SMMF) is the mean value of the line-of-sight (LOS) component of the solar vector magnetic field averaged over the visible hemisphere of the Sun.
So far, the studies on SMMF have mostly been confined to the magnetic field measurements at the photosphere.
In this study, we calculate and analyse the SMMF using magnetic field measurements at the chromosphere, in conjunction with that of photospheric measurements.
For this purpose, we have used full-disk LOS magnetograms derived from spectropolarimetric observations carried out in \FeIline and \CaIIline by the Synoptic Optical Long-term Investigations of the Sun (SOLIS)/Vector Spectromagnetograph (VSM) instrument during 2010 – 2017.
It is found from this study that the SMMF at the chromosphere is weaker by a factor of 0.60 compared to the SMMF at the upper-photosphere.
The correlation analysis between them gives a Pearson correlation coefficient of 0.80.
The similarity and reduced intensity of the chromospheric SMMF with respect to the photospheric SMMF corroborate the idea that it is the source of the Interplanetary Magnetic Field (IMF).

\end{abstract}


}]


\doinum{*}
\artcitid{\#\#\#\#}
\volnum{000}
\year{2022}
\pgrange{1--}
\setcounter{page}{1}
\lp{11}


\section{Introduction}

The Solar Mean Magnetic Field (SMMF) is the disk averaged value of the line-of-sight (LOS) magnetic field of the Sun.
It is also known as General Magnetic Field (GMF), Sun as a Star Magnetic Field (SSMF), Mean Magnetic Field (MMF) and global magnetic field, among others.
George Ellery Hale, who discovered magnetic fields on the Sun \citep{Hale1908}, was also the first to study the global magnetic field of the Sun \citep{Hale1913}.
A summary of the measurements of the global magnetic field of the Sun, carried out till the 1960s, is given in \citet{Severny1964}.

A number of studies have been done since then on the SMMF, particularly on the properties of the SMMF, its origin and its effects on the interplanetary magnetic field (IMF), which is the magnetic field in interplanetary space \citep{Wilcox-Ness1965, SEVERNY1969, Severny1970, Severny1971, Scherrer1972, Svalgaard1972, Svalgaard1975, Scherrer1977a, Kotov1983, Hoeksema1984, Sheeley1986, Demidov1998, Kotov1998, Demidov2002, Hudson2014, Sheeley-Wang2014, Sheeley2015, Sheeley2022}.
These studies have established the magnitude and periodicity of the SMMF.
They have shown that the SMMF has a non-zero value and a prominent periodicity of $\approx27$ days.
Its amplitude changes from $\approx\pm$\,0.2\,gauss during the solar minima to $\approx\pm$\,2\,gauss during the solar maxima \citep{Plachinda2011}.

On the other hand, a lot of uncertainty exists regarding the origin of the SMMF.
There is a possibility that the SMMF could be the remnant dipolar component of the primordial magnetic field present in the protostellar cloud during the formation of the Sun.
The argument for this goes as follows. \citet{Gough1998} established the necessity of having a large-scale magnetic field in the radiative zone of the Sun.
Later, \citet{Gough2017} provided supporting arguments for a part of this interior magnetic field to emerge to the surface, where we could observe it as the SMMF.
The very good correlation of the SMMF with the IMF augments this possibility.
On the other hand, \citet{Kutsenko2017} explained the SMMF as a result of the rotational modulation of high-intensity active region fields, and \citet{Ross2021} brought forward evidence that supported the connection of the rotationally modulated component of the SMMF to strong field regions like active regions and magnetic flux concentrations.
However, on the same topic, \citet{Bose2018} divided the solar surface into different features with the aid of image recognition and showed a high degree of correlation between the SMMF and the background solar field (and very little correlation between the SMMF and the field from active regions).

A comparison of the SMMF observed by different instruments was carried out by \citet{Demidov2000}.
Since then, comparisons of SMMF derived from the observations of different instruments (both space-borne and ground-based) have been done by many authors.
A good overview of these inter-instrument comparisons and references to earlier works are given in \citet{Pietarila2013} and \citet{Riley2014}.
These have largely been comparisons of disk-resolved magnetograms.
In \citet{Pietarila2013}, both disk-resolved and SMMF comparisons are made, and these two comparisons give significantly different results.

So far, all observations and analyses of the SMMF have been carried out using spectral lines formed at the photosphere.
In this paper, we use the \CaIIline spectral line to calculate the SMMF of the chromosphere and photosphere.
We validate our photospheric SMMF calculations by comparing them with available SMMF values from reference instruments and proceed to compare photospheric and chromospheric SMMF values.


\section{The Data}

\subsection{Data description}

The primary instrument used for data in this paper is the SOLIS - Vector Spectromagnetograph (VSM) \citep{Keller2001, Keller2003}.
SOLIS stands for Synoptic Optical Long-term Investigations of the Sun. It is a telescope facility with three instruments, VSM being one of them.
SOLIS has been offline since October 22, 2017, as it is being relocated and installed at the Big Bear Solar Observatory (BBSO), California.

\begin{figure*} 
  \centering
  \subfloat[]{\includegraphics[width=0.2\textwidth]{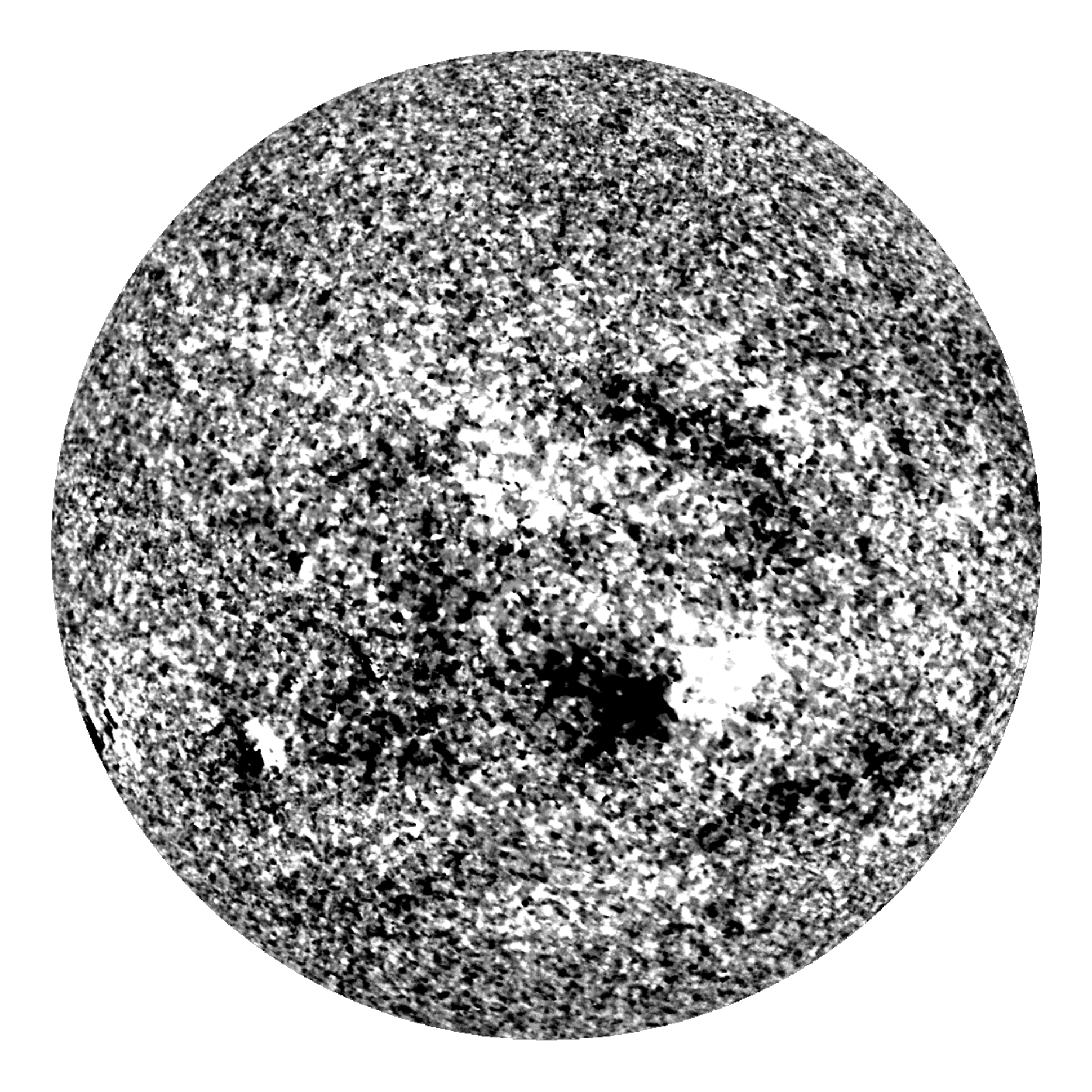}}
  \hspace{1em}
  \subfloat[]{\includegraphics[width=0.2\textwidth]{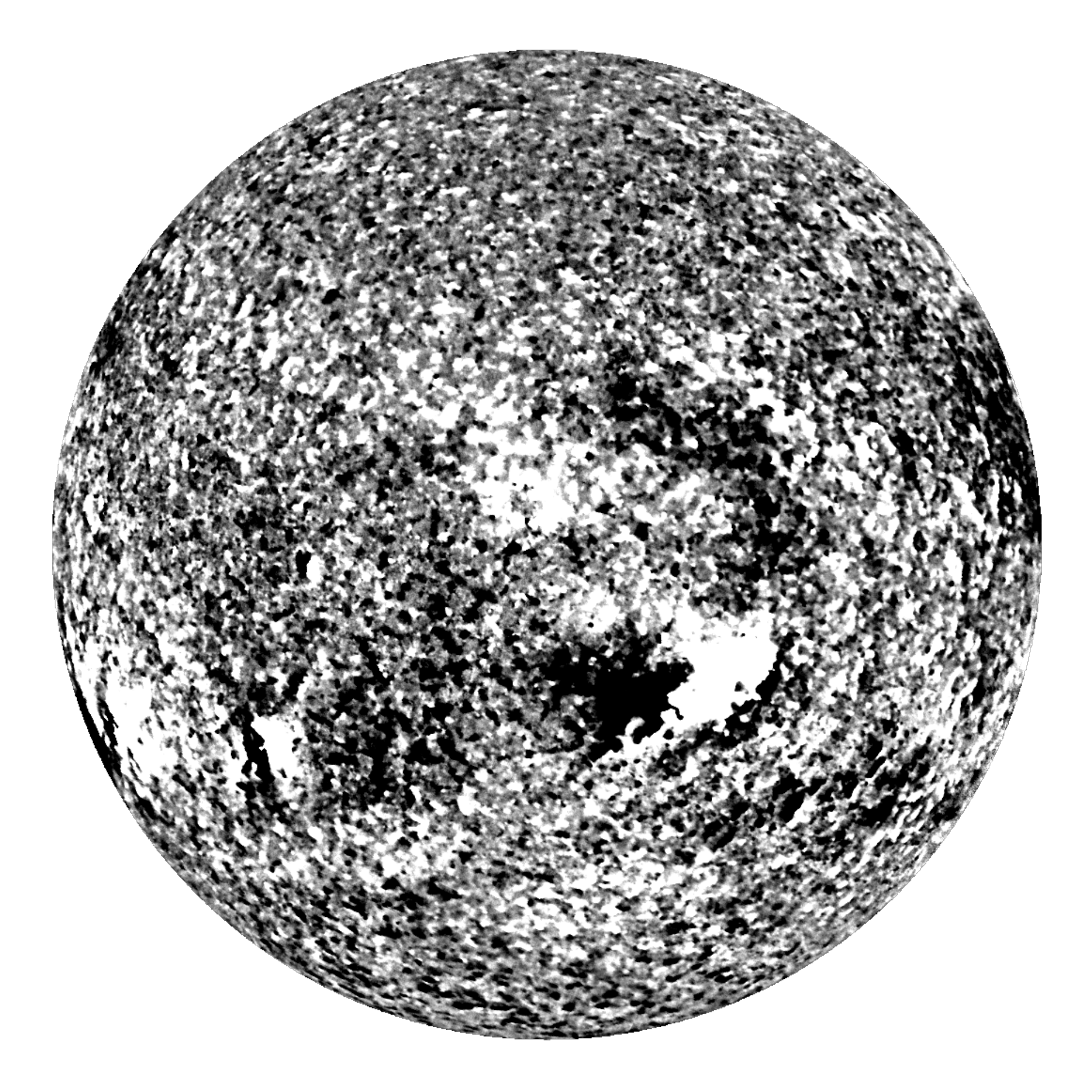}}
  \hspace{1em}
  \subfloat[]{\includegraphics[width=0.2\textwidth]{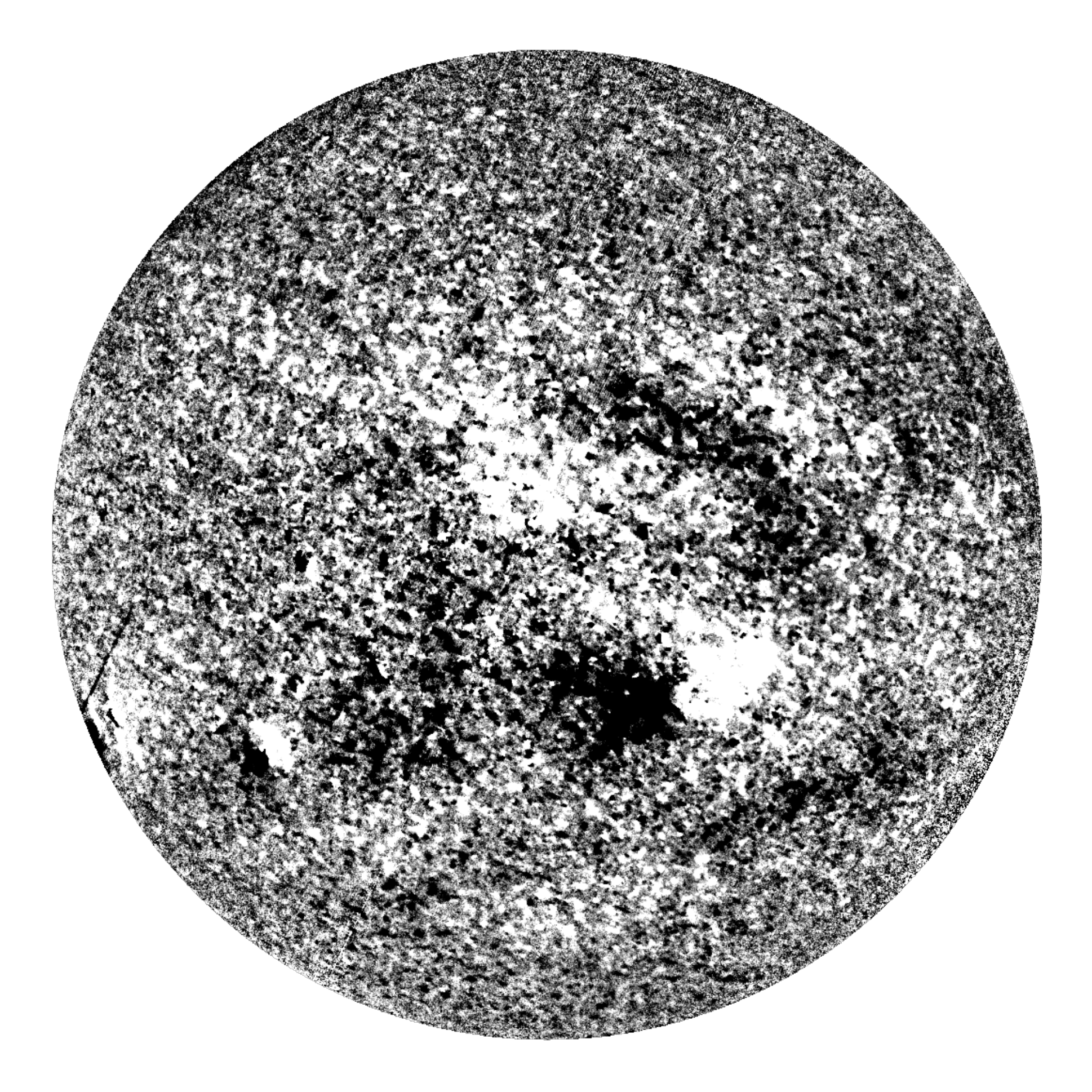}}
  \hspace{1em}
  \subfloat[]{\includegraphics[width=0.2\textwidth]{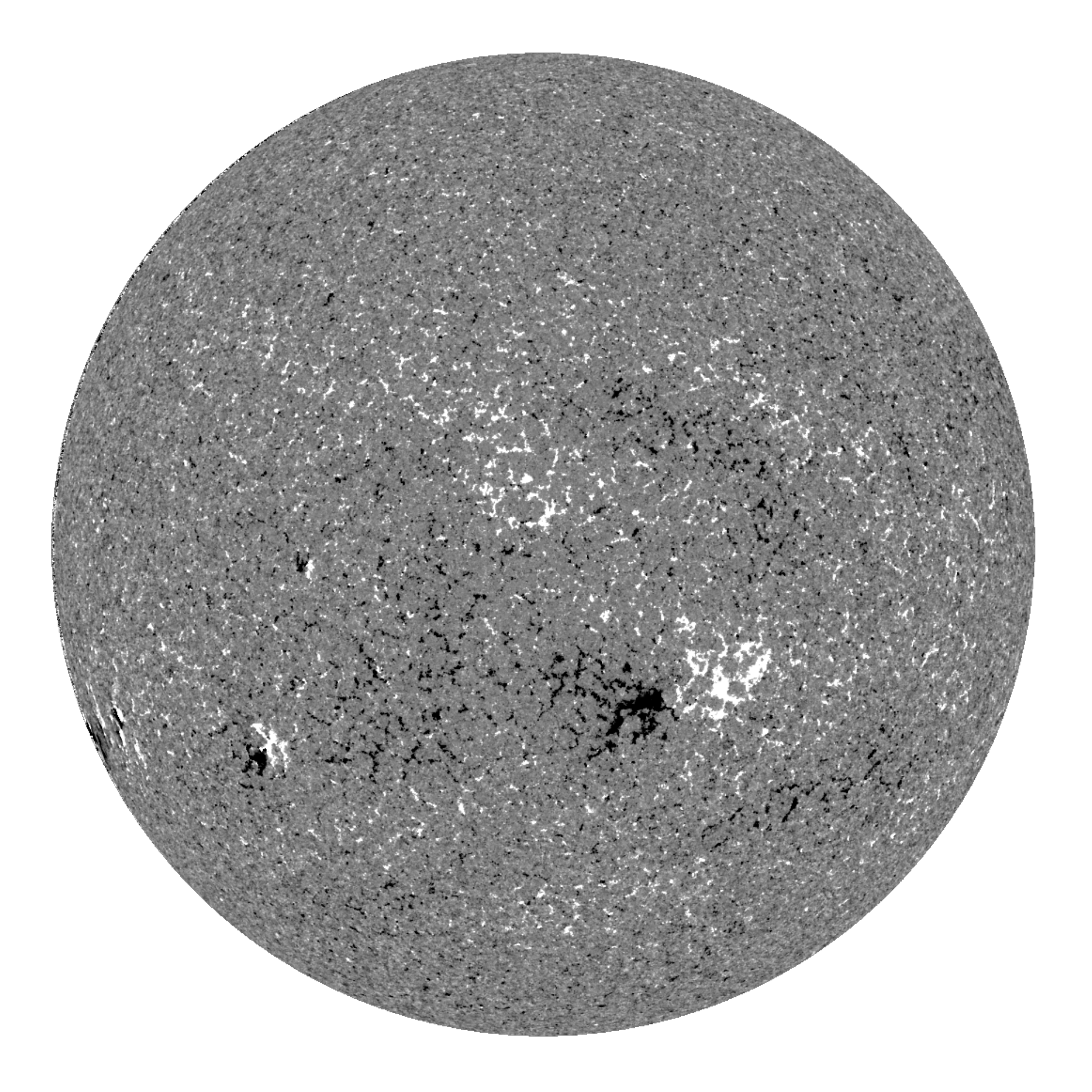}}
  \caption{(a) VSM 6302L magnetogram, (b) VSM 8542L magnetogram taken at the core of the spectral line (line-core data), (c) VSM 8542L magnetogram taken at the wings of the spectral line (line-wings data). (d) HMI 720s magnetogram. Observation date: 2015 October 8. Intensity thresholding is [-30, 30] gauss.\label{fig:vsmsamples}}
\end{figure*}

VSM is a Ritchey-Chr\'etien telescope fitted with a grating spectropolarimeter. \footnote{\url{https://nso.edu/wp-content/uploads/2018/04/VSM_details.pdf}}.
It has multiple modes of observation, among which we are interested in the 6302L and 8542L modes, which provide the full-disk, line-of-sight (LOS) magnetograms of the Sun at \FeIline and \CaIIline, respectively.
Samples of the 6302L and 8542L magnetograms are shown in Figure~\ref{fig:vsmsamples}.
Here, the 6302L data contains one magnetogram while the 8542L data contain two magnetograms; one generated using the core of the \CaIIline spectral line [8542.1\,\AA\,$\pm$\,600\,m\AA: hereafter called line-core data], and the other generated using the wings of the \CaIIline spectral line [8540.1\,\AA~-~8541.5\,\AA, 8542.7\,\AA~-~8544.1\,\AA: hereafter called line-wings data]\footnote{\label{note1}'SOLIS VSM 8542L full-disk images' at \url{https://nispdata.nso.edu/webProdDesc2/selector.php}}.
New CCD cameras added in January 2010 improved the spatial sampling of these magnetograms to 1\,$\mathrm{arcsec\,pixel^{-1}}$.
Using these magnetograms the SMMF values are calculated.

The 6302L mode takes 20 minutes to generate the magnetogram, while the 8542L mode takes 40 minutes for the process.
The magnetograms are generated daily, sometimes twice a day.
The observed LOS magnetograms, after relevant geometric and intensity corrections, are made available as Level 2 FITS files\footnote{\url{6302L data - ftp://solis.nso.edu/pubkeep/v72}, \url{8542L data - ftp://solis.nso.edu/pubkeep/v82}}.
We have used these files from 2010 May 1 to 2017 October 21 in our analyses.

The daily mean magnetic field data from the Wilcox Solar Observatory (WSO) at \wsoFeIline photospheric line is considered the reference for SMMF calculations.
This data is available from 1975 onwards\footnote{\url{http://wso.stanford.edu/\#MeanField}}.
\citet{Kutsenko2016} compared the WSO mean magnetic field to the SMMF calculated from full-disk magnetograms of the Helioseismic and Magnetic Imager (HMI) \citep{Scherrer2012} onboard the Solar Dynamics Observatory (SDO), and found a conversion factor of 0.99.
The HMI has better precision, faster cadence data than WSO.
This prompted us to use HMI SMMF measurements also to validate VSM SMMF values.
HMI is a filtergraph that observes the Sun at the \hmiFeIline photospheric line.
Among other data products, it provides full-disk LOS magnetograms at a cadence of 720 seconds.
A sample magnetogram is shown in Figure~\ref{fig:vsmsamples}.
We use the corresponding Level 1.5 science-ready data and the WSO mean field data, both taken during the same period as the VSM data, in our analyses.

We also use sunspot numbers in this paper to identify different stages of the solar cycle such as solar maximum, solar minimum, etc., and to study the variation of the SMMF with respect to the solar cycle.
Sunspot numbers are made available by the Sunspot Index and Long-term Solar Observations (SILSO), which is an activity under the Solar Influences Data analysis Center (SIDC), Royal Observatory of Belgium, Brussels.

\begin{table*}
\centering
\begin{tabular}{l l l l l}
 \hline\hline
  Instrument & Data & Spectral line & Method \\
 \hline
  WSO & Mean Field & \wsoFeIline & Zeeman splitting \\
  HMI & M\_720s (level 1.5) & \hmiFeIline & Doppler shift \\
  VSM & 6302L (level 2) & \FeIline & Zeeman splitting \\
  VSM & 8542L line-core (level 2) & \CaIIline & Zeeman splitting \\
  VSM & 8542L line-wings (level 2) & \CaIIline & Weak field approximation \\
 \hline
\end{tabular}
\caption{Instruments and spectral lines used to calculate SMMF in this work, and the method of calculation of magnetic field. \label{table:Sources_of_data}}
\end{table*}

\subsection{Details of magnetic field calculation and a review of comparison of data from different sources} \label{Field calculation details}

A summary of the sources and details of the various datasets used in this work are given in Table~\ref{table:Sources_of_data}.
The VSM spectrograph works in Littrow mode, and scans the solar disk from solar south to solar north, with its slit in the solar east-west direction \citep{Keller2003}.
In the 6302L mode, the magnetic flux density is calculated from Zeeman splitting, using a variant of the center-of-gravity method \citep{Jones2002}.
This mode has a spectral resolution of 23\,m\AA~\citep{Pietarila2013}.
In the 8542L mode, the magnetic flux density is calculated using the Weak Field Approximation (WFA)\footnote{'SOLIS VSM 8542L full-disk images' at \url{https://nispdata.nso.edu/webProdDesc2/selector.php}} \citep{spectral_lines_book}.
For this purpose, the spectral line is divided into bins of size 37.5\,m\AA, and the magnetic flux densities estimated from all bins are averaged\footnote{'SOLIS VSM 8542L full-disk images' at \url{https://nispdata.nso.edu/webProdDesc2/selector.php}}.
The use of WFA in the line-core introduces a systematic deviation in the measured magnetic field from the actual value at field intensities above $\approx$\,1200\,gauss \citep{Centeno2018}.
However, the pixels with magnetic field $>$\,1200\,gauss occupy a very small fraction of the solar disk, and calculating the SMMF cancels out this deviation to some extent.
This fraction was calculated to be $6.8\times10^{-6}$ for line-core data and $5.1\times10^{-5}$ for line-wings data in an 8542L magnetogram taken during solar maxima (2015/Feb/05).
The maximum value of instrumental noise is $\approx3\,\mathrm{gauss\,pixel^{-1}}$ in both modes \citep{Pietarila2013}.

The WSO uses a Babcock magnetograph with a grating spectrograph in the Littrow mode.
To obtain mean magnetic field data, the solar image is formed slightly above the spectrograph slit, thus feeding integrated light to the spectrograph.
The measurement error of the mean field is less than 0.05\,gauss \citep{Scherrer1977b}.
The LOS magnetic field is calculated using the Zeeman splitting of the \wsoFeIline photospheric line, from the change in intensity of the oppositely polarised parts of the spectral line, where the profile is steepest \citep{Babcock-Babcock1952, Beckers1968}.
Different authors give different multipliers for correcting the WSO magnetic field: $4.5-2.5\times\sin^2{\delta}$ \citep{Sheeley-Wang1995}, 1.85 \citep{Svalgaard2006}, no correction \citep{Riley2014}.
We have followed the latter and have not used any correction for the WSO data.

HMI uses a set of Lyot filters along with two Michelson interferometers to generate filtergrams.
It samples the spectral line at 6 wavelength points separated by 69\,m\AA.
The LOS magnetic field is calculated from the Doppler velocities obtained from these filtergrams at different polarisations.
The filter's effective FWHM is 76\,m\AA~and the random noise for this data is $\approx\,6.3\,\mathrm{gauss\,pixel^{-1}}$ \citep{Schou2012,Liu2012}.

\begin{figure*}[ht!] 
  \centering
  \includegraphics[width=1\textwidth]{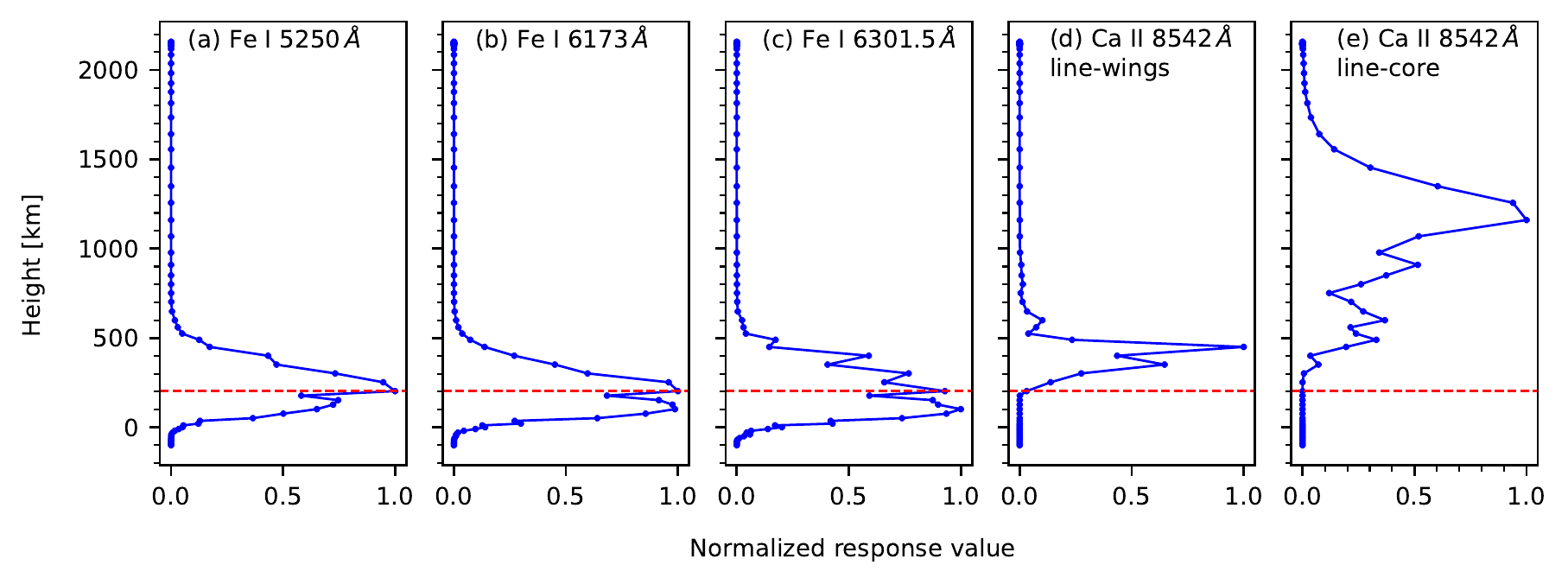}
  \caption{The response of the Stokes V profile to the perturbations in the LOS magnetic field, averaged over different magnetic fields, are presented with respect to height} for the spectral lines (a) WSO \wsoFeIline [5250.216\,\AA\,$\pm$\,100\,m\AA], (b) HMI \hmiFeIline [6173\,\AA\,$\pm$\,250\,m\AA], (c) VSM \FeIline [6301.515\,\AA\,$\pm$\,0.5\,\AA], (d) VSM \CaIIline line-wings [8540.1\,\AA - 8541.5\,\AA, 8542.7\,\AA - 8544.1\,\AA], (e) VSM \CaIIline line-core [8542.1\,\AA\,$\pm$\,600\,m\AA]. The functions are calculated at multiple points in the instrument’s wavelength range, and at the magnetic field values [50, 200, 500, 1000, 2000, 3000] gauss, and averaged. The red dotted line shows the peak of the response function for the WSO \wsoFeIline line.\label{fig:response_functions}
\end{figure*}

\citet{Pietarila2013} analyzed the magnetic flux densities measured from VSM and HMI magnetograms, and found that they are comparable.
They also calculated a linear fit of $MDI = -0.18 + 0.71 \times VSM$ by comparing the mean magnetic field between VSM and MDI.
Since MDI and HMI values are very much identical, our comparison of the HMI and VSM SMMF should give similar results.
\citet{Kutsenko2016} compared the WSO mean field and the disk-integrated magnetic field from HMI, and found a linear regression fit of $HMI = 0.99 \times WSO$ and a Pearson correlation coefficient of 0.86.
We would expect to get the same values from our analysis.
Similarly, inter-instrument comparisons of the magnetic field and SMMF at different spectral lines have been made by multiple authors.
So, despite the differences between instruments and spectral lines used for observations, we see that a comparison of the SMMF is quite plausible.
The heights of formation of these spectral lines are analyzed in the next subsection, to understand which pairs of spectral lines form at same heights in the solar atmosphere.

\subsection{Response functions of the spectral lines}

A response function provides the response of a spectral line to variations in atmospheric parameters.
Response functions are partial derivatives of a physical parameter ($X$) with height ($RF_{X}(h, \lambda) = \delta I(\lambda)/\delta X(h)$) \citep{Beckers1975}.
They are calculated by providing ``$+$'' and ``$-$'' perturbations ($\Delta x$) to the atmospheric parameter at each height and synthesising the spectral line profile, $S^{+}$ and $S^{-}$, respectively.
The response of the physical parameter at each height is then given by
\begin{equation}
RF_{X}(h, \lambda) = \frac{S^{+} - S^{-}}{2 * \Delta x}
\end{equation}

We calculated the response functions of the following lines in the context of our analyses: \wsoFeIline, \hmiFeIline, \FeIline and \CaIIline.
Please refer to these resources for similar calculations: \citet{QuinteroNoda2021, Rubio1997, QuinteroNoda2016}.
For our calculations, we used the atmospheric model FAL-C \citep{FAL1993} which represents the quiet Sun, and introduced magnetic fields of intensity [50, 200, 500, 1000, 2000, 3000] gauss.
The response of the Stokes V profile to the perturbations in the LOS magnetic field, averaged over the above magnetic fields, are presented with respect to height in Figure~\ref{fig:response_functions}.

We observe that the response functions of the \ion{Fe}{i} lines are very similar in shape.
Nevertheless, the \wsoFeIline line peaks at $\approx$200\,km above the surface, whereas the \FeIline line peaks at $\approx$100\,km above the surface.
The \hmiFeIline line has peaks at both $\approx$100\,km and $\approx$200\,km above the surface.
The plots also indicate that the \CaIIline line has contributions from both the photosphere and the chromosphere.
We observe that its line-core ($\pm$\,600\,m\AA~from the line-center) has contribution only from chromospheric heights (500\,km above the surface is the temperature minimum region in the FAL-C model), and its line-wings ($\pm$\,2\,\AA, excluding the line-core) have contribution only from upper photospheric heights.
Thus, we infer that the radiation from the wings of the \CaIIline line is originating from the upper photosphere, and the radiation from the core of the \CaIIline line is originating from the chromosphere.

\subsection{Data selection}

One magnetogram per day is considered for our analysis.
In general, the 6302L and 8542L data differ in their observation times.
We considered only the data for which both observations were taken within 2\,hours of each other.
This accounted for 88\% of all observations taken during the above stated period.
There was usually only one magnetogram available in a day, sometimes 2--3.
If more than one magnetogram satisfied the 2-hour condition, the earliest among them was selected.
If there were no magnetograms in a day that match the 2-hour condition, no datapoint was taken for that day.
There are 1506 days of observation satisfying this criterion, each day having one magnetogram each from 6302L, 8542L line-core, and 8542L line-wings data.

\begin{figure} 
  \centering
  \subfloat[\label{fig:badpixels}]{
  \includegraphics[width=0.2\textwidth]{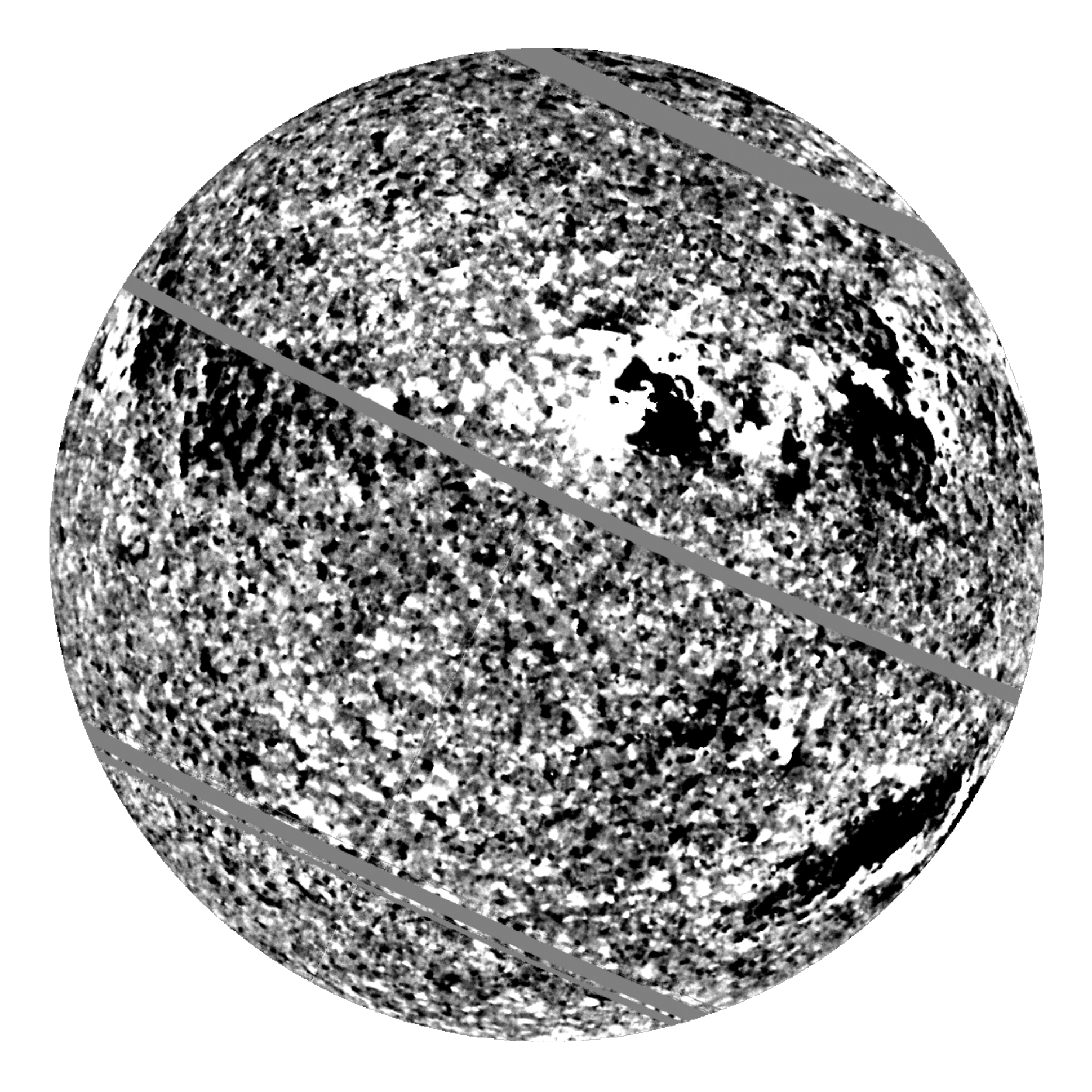}
  \hspace{1em}
  \includegraphics[width=0.2\textwidth]{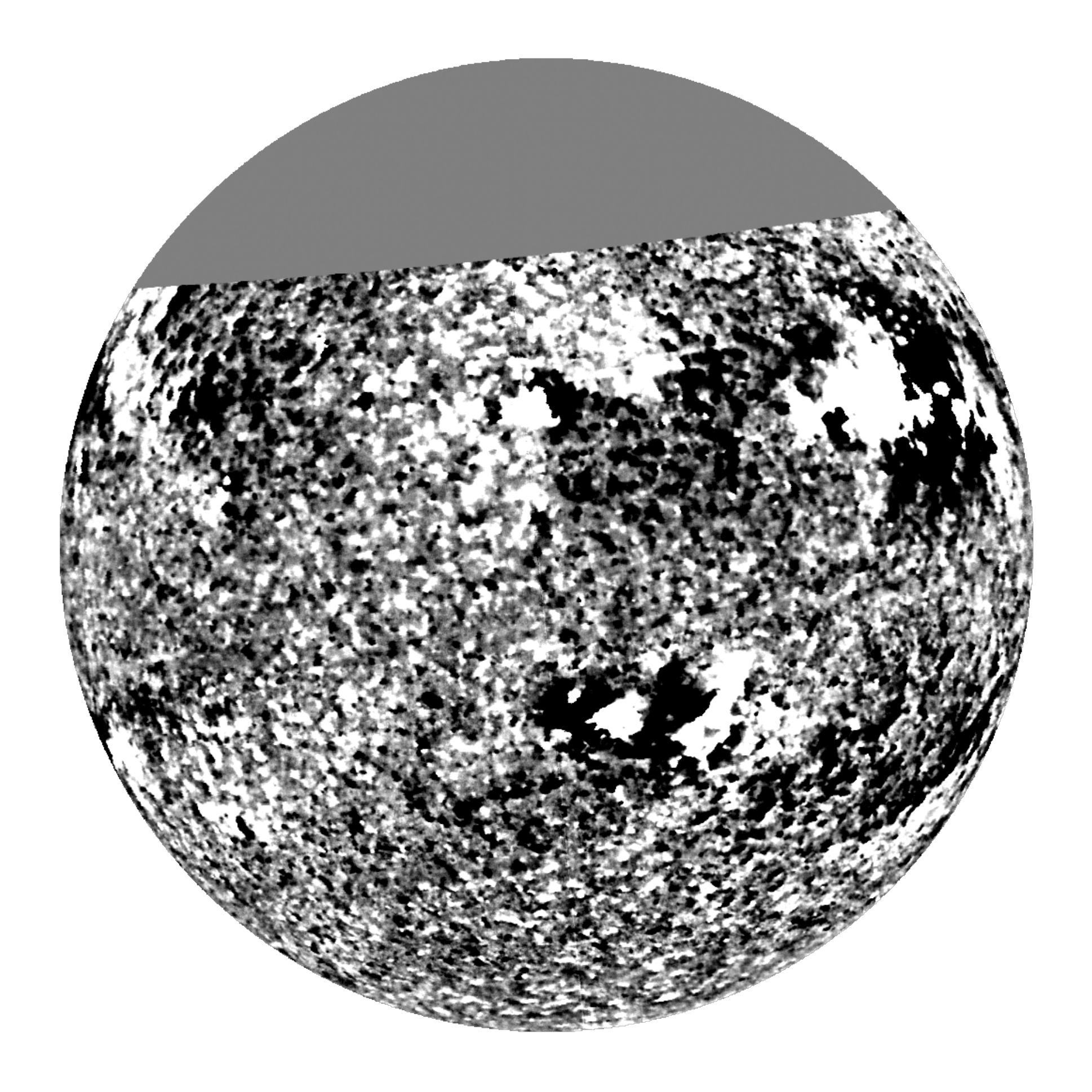}}
  
  \subfloat[\label{fig:stripes}]{\includegraphics[width=0.2\textwidth]{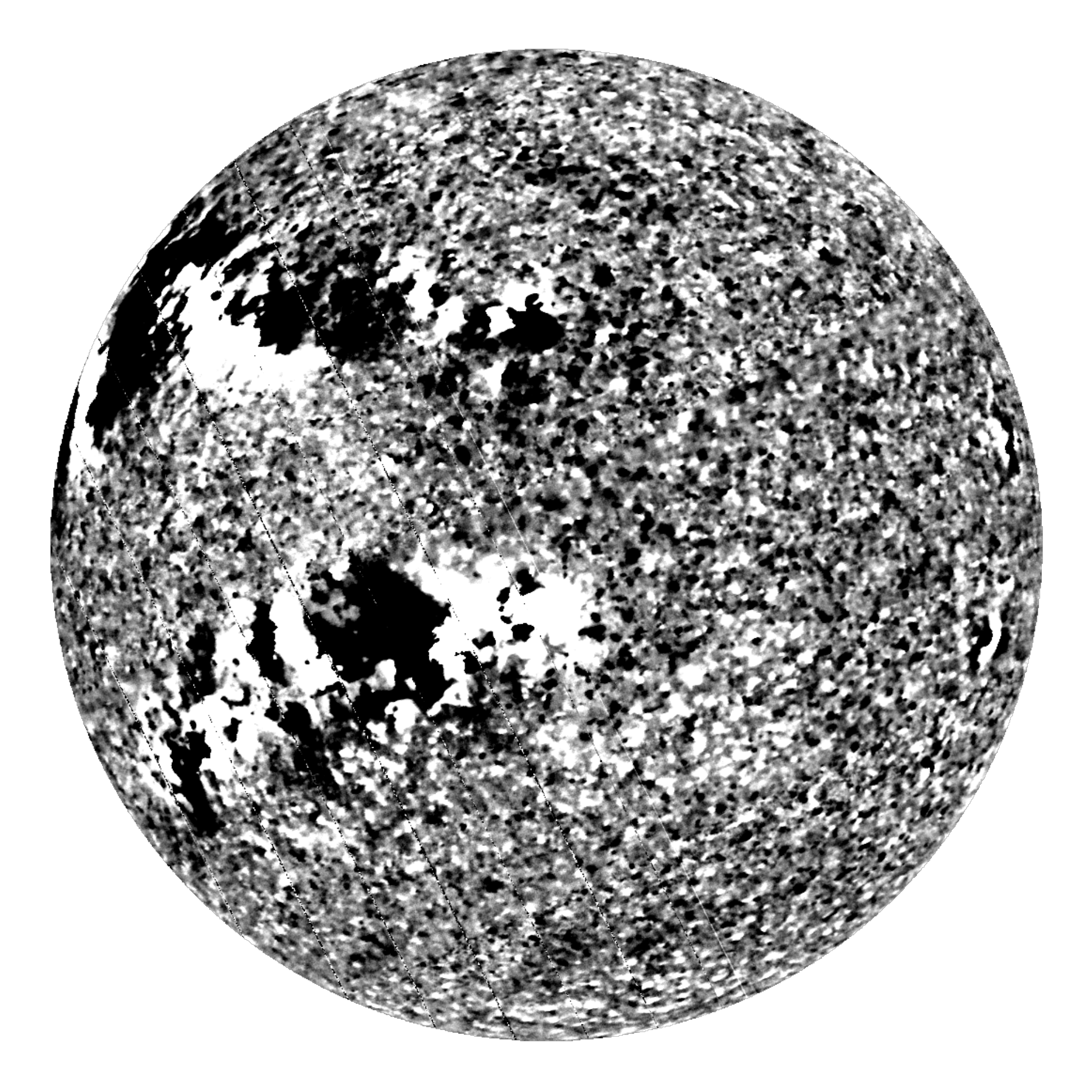}}
  \caption{(a) Sample VSM magnetograms that have many bad pixels. Intensity thresholding is [-5, 5] gauss. (b) A sample VSM magnetogram that has artifacts shaped as stripes. Intensity thresholding is [-60, 30] gauss.}
\end{figure}

It was observed that there are pixels in the VSM magnetograms having values of {\itshape Not-A-Number (nan)} or {\itshape zero}.
These could have come during the acquisition or processing of the data.
Unaffected data always has a floating point value. We termed such pixels containing either {\itshape nan} or {\itshape zero} values as bad pixels.
Sample magnetograms with bad pixels are shown in Figure~\ref{fig:badpixels}.
We observed that many of the bad pixels appeared as bands of missing data in magnetograms and that the width of these bands increased as the fraction of bad pixels in the solar disk increased.
We decided to discard magnetograms with bands of size $\ge$\,20\,arcseconds. This corresponded to 0.45\% or more of the solar disk covered by bad pixels.
When one magnetogram among the three modes (6302L, 8542L line-core, 8542 line-wings) was discarded, the others were discarded too.
These corresponded to 41 days of data (datapoints).
Apart from the above bad pixels, some magnetograms were observed to have artifacts shaped as stripes across the solar disk (Figure~\ref{fig:stripes}).
These magnetograms were also discarded, leaving 1459 datapoints.

Once the magnetograms were selected, for each magnetogram (say 6302L), the bad pixels were located and {\itshape nan} value was assigned to the pixels at the same locations in its complementary magnetograms (8542L line-core and line-wings).
This was repeated for each of the 8542L magnetograms.
This makes sure that the magnetograms are comparable.
All magnetograms (6302L, 8542L line-core, and 8542L line-wings) are already North-corrected in Level 2.
The solar center changes by less than a pixel between corresponding 6302L/8542L files; this difference is neglected.
For each magnetogram, a circular mask centered at the solar center and with a radius equal to 99.5\% of the solar radius was generated to select the solar disk.
The solar disk boundary was avoided because of the sharp change in intensity.
The values of the solar center and solar radius were read from the corresponding FITS header.
The arithmetic mean of all pixels within this mask was calculated and recorded as the Solar Mean Magnetic Field (SMMF).

Similarly, for each HMI magnetogram a mask was created and the arithmetic mean of the pixels within this mask was recorded as the SMMF.
We used two parameters in the header file to select HMI data - QUALITY and QUALLEV1.
They are non-zero for calibration data, and also when the data is not good.
We excluded all data with non-zero values of either of these two parameters, and random checking of the selected magnetograms showed that \textit{nan} values were absent in the solar disk.
1267 magnetograms were selected by this process, which were all taken within 2\,hours of VSM observations.

We define SMMF (6302L) as \SMMFphotosphere, SMMF (8542L line-wings) as \SMMFwings, SMMF (8542L line-core) as \SMMFcore, SMMF (WSO) as \SMMFwso and SMMF (HMI) as \SMMFhmi.
The average observation times of corresponding 6302L and 8542L datapoints were calculated, and the SMMF values were associated with them to facilitate comparisons.

\subsection{Data validation}

The \textit{VSM photospheric mean net flux} is a dataset present in the SOLIS website\footnote{\url{https://solis.nso.edu/0/vsm/vsm\_mnfield.html}}.
Each datapoint in this dataset corresponds to a magnetogram, and has the average value of all pixels within 99\% of the solar disk.
\SMMFphotosphere compares well with this mean net flux. Choosing datapoints observed at the same time (since the latter could have multiple datapoints/day), we get 1489 common datapoints between them, which yield a linear regression coefficient ($\beta$) of 0.99 (\SMMFphotosphere v/s net flux) and a correlation coefficient ($\gamma$) of 0.99.
This confirms the accuracy of our calculation of \SMMFphotosphere.

Similarly, each of the HMI magnetograms' FITS header file contains a parameter called DATAMEAN, which is equal to the disk-averaged flux.
We compared \SMMFhmi with the value of the DATAMEAN parameter.
They matched very well, with $\beta = 0.98$ (\SMMFhmi v/s DATAMEAN) and $\gamma = 1.00$.
This corroborates our calculation of \SMMFhmi.

However, we have not used either the VSM photospheric mean net flux or the HMI DATAMEAN parameter as the SMMF.
This is because, the DATAMEAN parameter is calculated for all magnetograms, whether they are calibration files, observation files or erroneous files, and the VSM photospheric mean net flux is calculated for magnetograms with erroneous data (artifacts) also.
As mentioned in the previous subsection, we have discarded such erroneous data.

\subsection{Data analysis}

\begin{figure*} 
  \centering
    \includegraphics[width=1\textwidth]{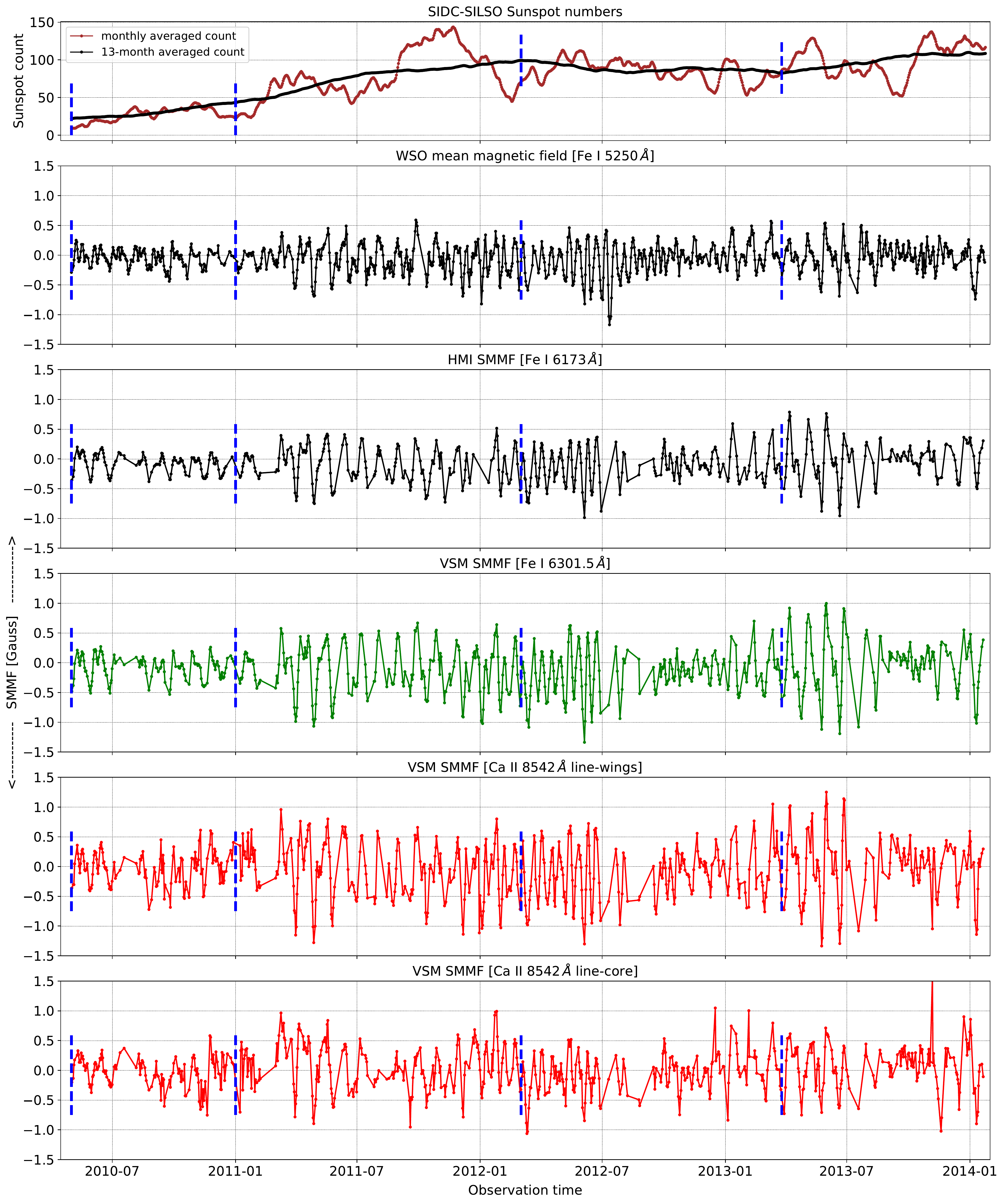}
  \caption{The sunspot numbers (monthly averaged and 13-months averaged), daily \SMMFwso, \SMMFhmi and VSM SMMF plots from 2010/05/01 to 2014/01/24. \SMMFphotosphere, \SMMFcore, and \SMMFwings are plotted according to the average observation time between them. Each datapoint is shown by a dot, whereas lines connect adjacent datapoints. Data gaps can be understood by the absence of dots in the line. The vertical blue lines demarcate different stages of the solar cycle, namely, solar minimum, rising phase, solar maximum, and declining phase}.\label{fig:compreplot1}
\end{figure*}

\begin{figure*} 
  \centering
    \includegraphics[width=1\textwidth]{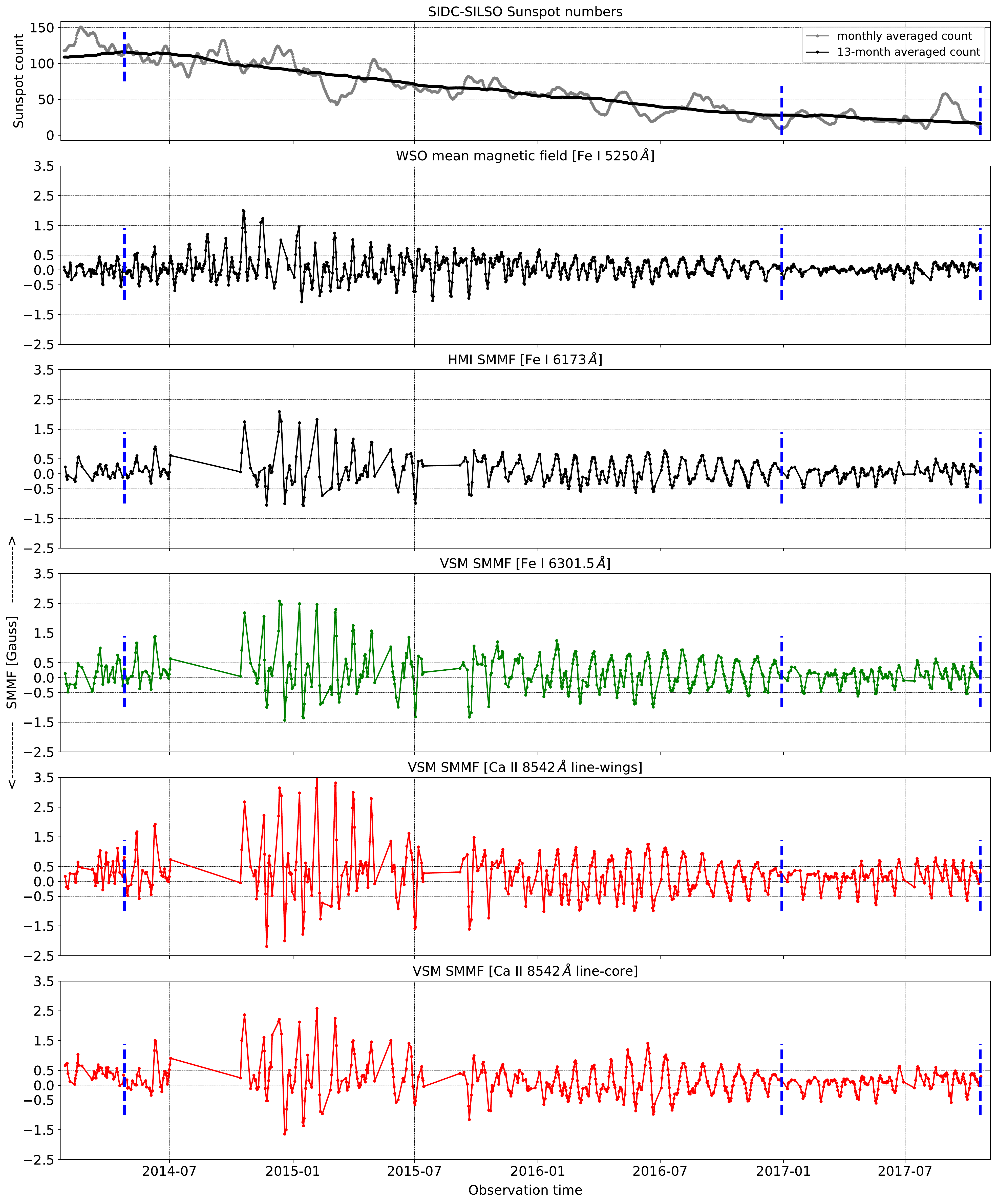}
  \caption{The sunspot numbers (monthly averaged and 13-months averaged), daily \SMMFwso, \SMMFhmi and VSM SMMF plots from 2014/01/25 to 2017/10/21. \SMMFphotosphere, \SMMFcore, and \SMMFwings are plotted according to the average observation time between them. Each datapoint is shown by a dot, whereas lines connect adjacent datapoints. Data gaps can be understood by the absence of dots in the line. The vertical blue lines demarcate different stages of the solar cycle, namely, solar minimum, rising phase, solar maximum, and declining phase}.\label{fig:compreplot2}
\end{figure*}

The plots of daily VSM SMMF, \SMMFwso, \SMMFhmi, and sunspot numbers (13-month running average, and monthly running average) are shown in Figures~\ref{fig:compreplot1}, \ref{fig:compreplot2}.
\SMMFphotosphere, \SMMFcore and \SMMFwings are plotted according to their average observation times.
Each datapoint is shown by a dot, whereas lines connect adjacent datapoints.
Data gaps can be understood by the absence of dots in the line.
The VSM data has 1267 common datapoints with \SMMFhmi, and 1241 common datapoints with \SMMFwso.
\SMMFphotosphere, \SMMFwings and \SMMFcore have 1459 common datapoints.
The vertical blue lines demarcate different stages of the solar sunspot cycle, such as solar minimum, rising phase, solar maximum, and declining phase.
This has been done using the 13-month running averaged sunspot number from SILSO.

\begin{figure}[t!] 
  \includegraphics[width=0.485\textwidth]{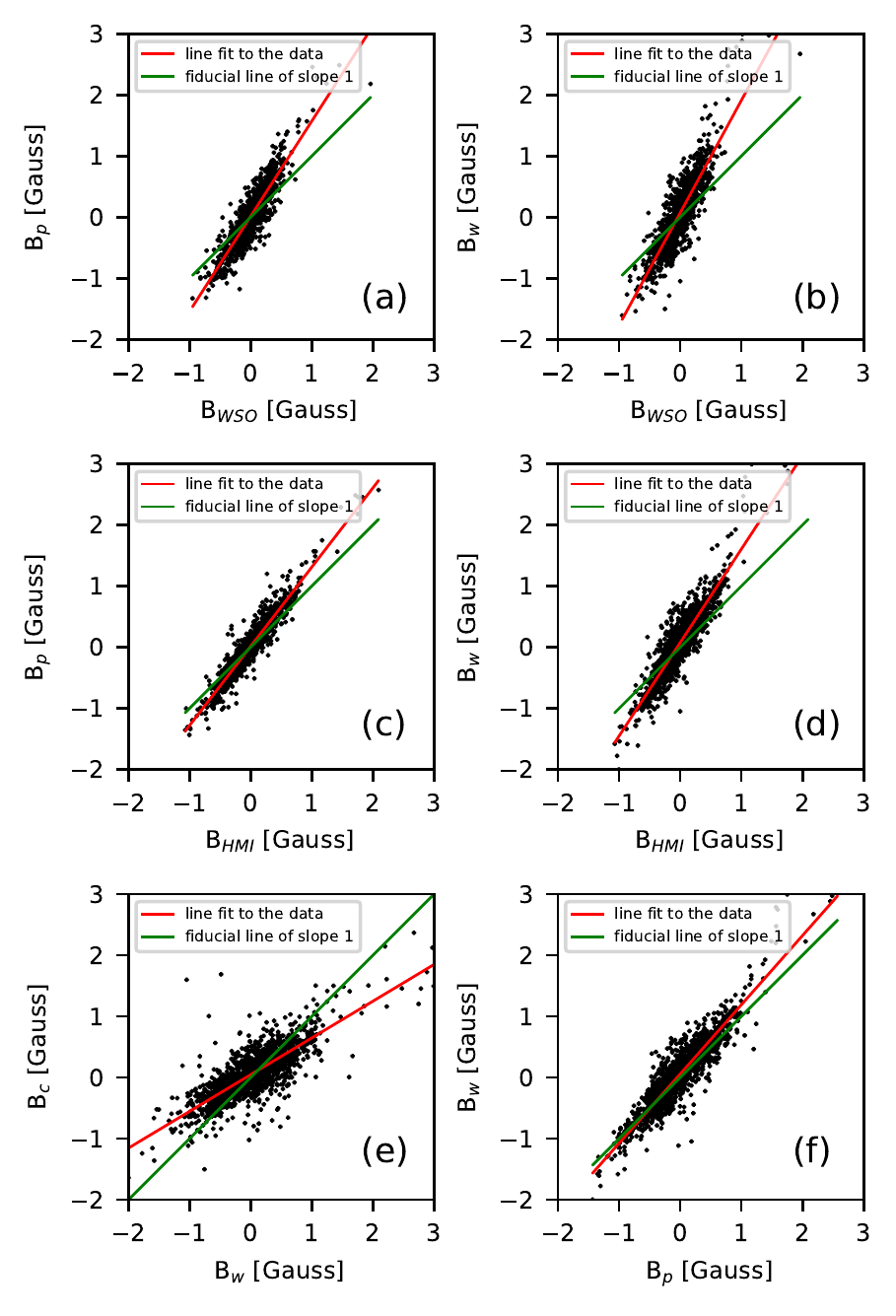}
  \caption{Scatter plots of: (a) \SMMFphotosphere v/s \SMMFwso, (b) \SMMFwings v/s \SMMFwso, (c) \SMMFphotosphere v/s \SMMFhmi, (d) \SMMFwings v/s \SMMFhmi, (e) \SMMFcore v/s \SMMFwings, (f) \SMMFwings v/s \SMMFphotosphere. \label{fig:scatterplots}}
\end{figure}

A visual inspection of Figures~\ref{fig:compreplot1}, \ref{fig:compreplot2} show that VSM SMMFs, \SMMFwso, and \SMMFhmi are similar in shape, but have differences in their amplitudes.
Correlation and regression coefficients between the SMMF data for the complete observation time are presented in Table~\ref{table:CorrCoeff}, and the scatter plots for the same are given in Figure \ref{fig:scatterplots}.
We also calculated these coefficients within each stage of the sunspot cycle (mentioned above), but did not observe any patterns.
They were also calculated according to the 13-month running averaged sunspot numbers, during periods of low solar activity (count $<$ 50, 40, 30, 20), and high solar activity (count $>$ 50, 60, 70, 80, 90, 100, 110).
These results are mentioned in Table~\ref{table:CorrCoeff_corenwings}.
From this table, we observe that the correlation between \SMMFwings and \SMMFcore is higher at periods of low solar activity (lower sunspot numbers) compared to periods of high solar activity (higher sunspot numbers).

\begin{table}
\begin{center}
\begin{tabular}{c c c c c}
 \hline\hline
  & \multicolumn{4}{c}{$\gamma$} \\
 \hline
  & \SMMFphotosphere & \SMMFwings & \SMMFcore & \SMMFhmi \\
 \hline
  \SMMFwso & 0.89 & 0.87 & - & 0.86 \\
  \SMMFhmi & 0.95 & 0.92 & - & - \\
  \SMMFphotosphere & - & 0.92 & - & - \\
  \SMMFwings & - & - & 0.80 & - \\
 \hline\hline
  & \multicolumn{4}{c}{$\beta$} \\
 \hline
  & \SMMFphotosphere & \SMMFwings & \SMMFcore & \SMMFhmi \\
 \hline
  \SMMFwso & 1.56 & 1.84 & - & 1.09 \\
  \SMMFhmi & 1.30 & 1.53 & - & - \\
  \SMMFphotosphere & - & 1.13 & - & - \\
  \SMMFwings & - & - & 0.60 & - \\
 \hline\hline
\end{tabular}
\caption{Correlation coefficients ($\gamma$) and linear regression coefficients ($\beta$) between \SMMFwso, \SMMFhmi, \SMMFphotosphere, \SMMFcore and \SMMFwings, calculated over the entire observation period from 2010 to 2017. How to read the table: for the linear regression coefficient, the parameter in the first column is the independent variable, and the parameter in the row is the dependent variable. \label{table:CorrCoeff}}
\end{center}
\end{table}

\begin{table}[b!]
\begin{center}
\begin{tabular}{c c c c c c c c}
 \hline\hline
  & \multicolumn{6}{c}{Sunspot number} \\
 \hline
  & $< 20$ & $< 30$ & $< 40$ & $< 50$ & $> 50$ & $> 60$ \\
 \hline
  $\gamma$ & 0.82 & 0.84 & 0.85 & 0.83 & 0.80 & 0.80 \\
 \hline\hline
  & $> 70$ & $> 80$ & $> 90$ & $> 100$ & $> 110$ \\
 \hline
  $\gamma$ & 0.80 & 0.79 & 0.74 & 0.49 & 0.57 \\
 \hline
\end{tabular}
\caption{Correlation coefficients ($\gamma$) between \SMMFcore and \SMMFwings calculated over different periods corresponding to the number of visible sunspots. How to read the table: sunspot number $> 70$ implies that the correlation was calculated for only those time periods during which the number of sunspots was $> 70$. \label{table:CorrCoeff_corenwings}}
\end{center}
\end{table}

We also compared \SMMFhmi with \SMMFwso during the entire VSM observation period (1115 datapoints), and found $\gamma = 0.86$ and $\beta = 1.09$ (HMI v/s WSO).
We note that the value of $\gamma$ is exactly same as that of \citet{Kutsenko2016}, while there is a slight difference in the value of $\beta$; this could be because of a few reasons.
\citet{Kutsenko2016} had used 1507 data pairs between January 1, 2011 and December 10, 2015 in their analysis.
So, one possibility is that the two datasets were taken at different times.
Another possibility is that the former authors had included HMI magnetograms with non-zero values of the QUALITY parameter in their analysis if \textit{``the magnetogram showed no abnormal value of the SMMF"}.
Also, we have used the HMI level 1.5 data, and although we assume that the former authors have used the same level of data, it's not mentioned in their paper.


\section{Results and Discussion}

We have used both \SMMFwso and \SMMFhmi as references in our calculations.
As mentioned in section 2, the HMI data has better precision compared to WSO.
We can notice this in Figure~\ref{fig:compreplot2}, where periodicities during the first half of 2017 are hard to notice in \SMMFwso, whereas visible in \SMMFhmi.
But, \SMMFhmi data is only available from 2010 May 1 onwards.
Thus, we have used both \SMMFwso and \SMMFhmi in this study.
We have tabulated the correlation ($\gamma$) and linear regression ($\beta$) coefficients between the VSM SMMF datasets and the reference datasets (\SMMFwso and \SMMFhmi), and also within the VSM SMMF datasets.
The response functions of VSM \FeIline line, WSO \wsoFeIline line, and HMI \hmiFeIline line are similar in shape, i. e., these spectral lines have similar heights of formation, and similar contributions from each height of the solar atmosphere.
Thus, they probe a similar region in the solar atmosphere.
This, coupled with the good correlation of the associated SMMF datasets, leads us to consider $\beta$ to be the scaling factor between these datasets.
The scaling factors are as follows: \SMMFphotosphere $= 1.56 \times $\SMMFwso, \SMMFphotosphere $= 1.30 \times $\SMMFhmi.
However, the response function of the VSM \CaIIline line-wings peaks at a different height and has different contributions from atmospheric heights compared to the former response functions.
Thus, $\beta$ cannot be considered as the scaling factor between these datasets.

\SMMFcore and \SMMFwings have $\beta = 0.60$.
Since both measurements are taken from the same instrument using the same procedure, no further calibration is needed.
It directly gives the conversion factor of chromospheric SMMF to photospheric SMMF.
$\beta < 1$ implies a reduction in intensity from the photospheric SMMF to the chromospheric SMMF.
We also note that $\gamma = 0.80$ between \SMMFwings and \SMMFcore, and that it is higher at periods of low sunspot numbers as compared to periods of high sunspot numbers.
But, as mentioned in section \ref{Field calculation details}, the WFA underestimates the magnetic field at high field intensities.
As helpfully mentioned by our referees, this could result in improved correlation at periods of low magnetic field (low sunspot numbers).
To verify this possibility, we calculated the correlation coefficients between the SMMF datasets using pixels with magnetic field $< 1200$\,gauss, and found that there is no effect of WFA on the correlation.
A drawback in this study is that since \SMMFwings has contribution only from upper photospheric heights, we are not having information from all heights of the photosphere.
This is related to the wavelength range (600\,m\AA~-~2\,\AA~ from the center of the spectral line) which is available as \CaIIline line-wings data.
So, a future work could be to consider different wavelength ranges (further away from the line-core), check their Stokes-V response functions, and select the one which has a contribution from all photospheric heights.


\section{Conclusion}

In this paper, we have calculated and compared the photospheric (\SMMFphotosphere and \SMMFwings) and chromospheric (\SMMFcore) SMMF values estimated from VSM full-disk magnetograms.
We used the mean field data from WSO (\SMMFwso) and HMI (\SMMFhmi) as references to validate the VSM SMMF values and observed that the SMMF values between different instruments retain similar features but differ in their magnitudes.
We note that the difference in values between VSM and WSO/HMI could be because of the instrument conversion factor.
On the other hand, the comparison between VSM 8542 line-core (chromospheric SMMF) and line-wings (upper photospheric SMMF) does not contain any effects of the instrument or measurement techniques, and their linear regression coefficient can be considered as the ratio of the SMMF at these heights.
It was found that the chromospheric SMMF is weaker by a factor of 0.6 compared to that of upper photospheric SMMF.
This reduction in intensity could mean that a significant part of the SMMF is a magnetic field that propagates outwards from the photosphere to the chromosphere.
This is in line with the view that the SMMF could have a source, decoupled from solar activity.
It was also found that the correlation between \SMMFcore and \SMMFwings is higher during periods of lower solar activity.
The similarity between the photospheric and chromospheric SMMFs, and the reduced intensity of the chromospheric SMMF with respect to the photospheric SMMF corroborate the idea that the SMMF could be passing through the photosphere, chromosphere, and arriving at interplanetary space, where we measure it as the IMF.


\section*{Acknowledgements}

We acknowledge the data centers: VSM/SOLIS, NSO; WSO, Stanford; HMI/SDO; WDC-SILSO, Royal Observatory of Belgium, Brussels.

This research used version 3.1.6 \citep{sunpy_community2020} of the SunPy open source software package \citep{Barnes_2020}. We have also used the packages h5py \citep{collette_python_hdf5_2014}, matplotlib \citep{Hunter2007}, numpy \citep{harris2020array} and scipy \citep{2020SciPy-NMeth} to carry out our data analysis.


\bibliography{refs}

\end{document}